\documentclass[a4paper]{article}

\usepackage{INTERSPEECH2022}

\title{Production federated keyword spotting via distillation, filtering, and joint federated-centralized training}
\name{Andrew Hard$^{1}$, Kurt Partridge$^{1}$, Neng Chen$^{1}$, Sean Augenstein$^{1}$, Aishanee Shah$^{1}$, Hyun Jin Park$^{1}$, Alex~Park$^{1}$, Sara~Ng$^{1,2}$, Jessica Nguyen$^{1}$, Ignacio Lopez Moreno$^{1}$, Rajiv Mathews$^{1}$, Fran{\c{c}}oise~Beaufays$^{1}$}

\address{
  $^1$Google LLC, Mountain View, CA, U.S.A. \\
  $^2$University of Washington, Seattle, WA, U.S.A.
}
\email{\{harda,kep,nengchen,saugenst,aishaneeshah,hjpark,\\axpk,sbng,jessnguyen,elnota,mathews,fsb\}@google.com}

\begin{document}

\maketitle

\begin{abstract}
  We trained a keyword spotting model using federated learning on real user
  devices and observed significant improvements when the model was deployed for
  inference on phones. To compensate for data domains that are missing from
  on-device training caches, we employed joint federated-centralized training.
  And to learn in the absence of curated labels on-device, we formulated a
  confidence filtering strategy based on user-feedback signals for federated
  distillation. These techniques created models that significantly improved
  quality metrics in offline evaluations and user-experience metrics in live
  A/B experiments.
\end{abstract}
\noindent\textbf{Index Terms}: federated learning, keyword spotting, filtering,
distillation, multi-task learning, domain adaptation

\section{Introduction}
\label{sec:introduction}

Keyword spotting (KWS) models provide an important access point for virtual
assistants. \textit{Alexa}, \textit{Hey Google}, and \textit{Hey Siri} are just
a few examples of keywords that can be used to initiate queries and issue
commands to various phone and smart speaker devices. The underlying algorithms
must efficiently process streaming audio to trigger on instances of the keyword
while screening out the vast majority of irrelevant noise.

Neural networks have been thoroughly studied in the context of KWS. Prior works
have shown significant quality improvements and latency reductions for
low-resource inference settings~\cite{small_fp_kws, mtl_wce_kws, hey_siri,
comp_td_nn_kws, MaxPool20, cascade_kws}.

Federated learning (FL)~\cite{fedlearn} is a privacy-enhancing distributed
computation technology that can be used to improve neural models on-device
without sending users' raw data to servers. Model updates are communicated from
clients to the server, but they are ephemeral (only stored in temporary memory),
focused (only relevant to the training task at hand), and are only released in
aggregate.

Recently, FL has been applied to many production domains including keyboard
next-word prediction~\cite{flnwp} and speaker
verification~\cite{Granqvist2020ImprovingOS}. Prior works have also used FL
algorithms to train KWS models in simulation
environments~\cite{fed_kws, FedHotNonIID20}. In most cases, federated training
benefits models as a result of the abundant in-domain data available on-device.

Even so, there is often a mismatch between the domain of the cached data for FL
and the inference data domain: certain device types or conditions may be
underrepresented. In such cases, it is beneficial to combine FL with
server-based training. One approach, federated transfer
learning~\cite{Liu2018SecureFT}, suffers from catastrophic
forgetting~\cite{FRENCH1999128, Ratcliff1990ConnectionistMO}. Another more
recent approach is joint training, in which server-based learning is combined
with FL~\cite{augenstein2021mixedfl}.

While supervised learning has been the dominant approach to KWS model training,
semi-supervised~\cite{Oord2018RepresentationLW, Schneider2019wav2vecUP,
Zhang2021BigSSLET} and self-supervised~\cite{e2e2e_neurips2019, nst_img_2020}
training techniques have become popular as they are able to leverage large
unlabeled datasets to achieve comparable or superior performance to supervised
models. These techniques are particularly helpful in the cases  of FL (where
the client cache data are ``no-peek" by design and cannot be manually labeled)
and speech (where audio is abundant but the labeling process is arduous).
Distillation~\cite{distillation} of a teacher model into a student with
equivalent architecture has also been used previously for KWS~\cite{KwsStce21}.

Filtering noisy labels has a long history and can be seen as a form of
curriculum learning~\cite{10.1145/1553374.1553380}.  More recently it has been
explored in semi-supervised settings in ASR~\cite{hwang2021large}.
See~\cite{han2020survey,song2020learning} for more detailed surveys. However,
this literature does not address how to infer label confidence using additional
correlated features that are only available at training time.

This paper presents a successful application of federated learning for the
speech keyword spotting domain. Several unique contributions are provided,
including joint federated-centralized training and on-device example filtering
based on feedback feature analysis.

\section{Keyword spotting model}
\label{sec:model}

\subsection{Input features}

The input features used in this work were identical to those used in prior
publications~\cite{MaxPool20, Alvarez2019}. A 40-dimensional vector of spectral
filter-bank energies was extracted at each time frame $t$ (generated every 10ms
over a 25ms window). 3 temporally-adjacent frames were stacked and strided to
produce a 120-dimensional input feature vector, $X_t$, every 20ms.

To improve model robustness and generalization, data augmentation techniques
were applied to the features. During server training, classic augmentation
methods such adding simulated reverberation and mixing noise were applied to
the data prior to feature extraction, as in \cite{kim2017generation}. These
methods were infeasible for on-device training~\cite{FedHotNonIID20}, so
spectral augmentation~\cite{SpecAug19} was used instead with FL.

\subsection{Architecture}
\label{sec:architecture}

\begin{figure}
	\centering
	\includegraphics[width=\columnwidth]{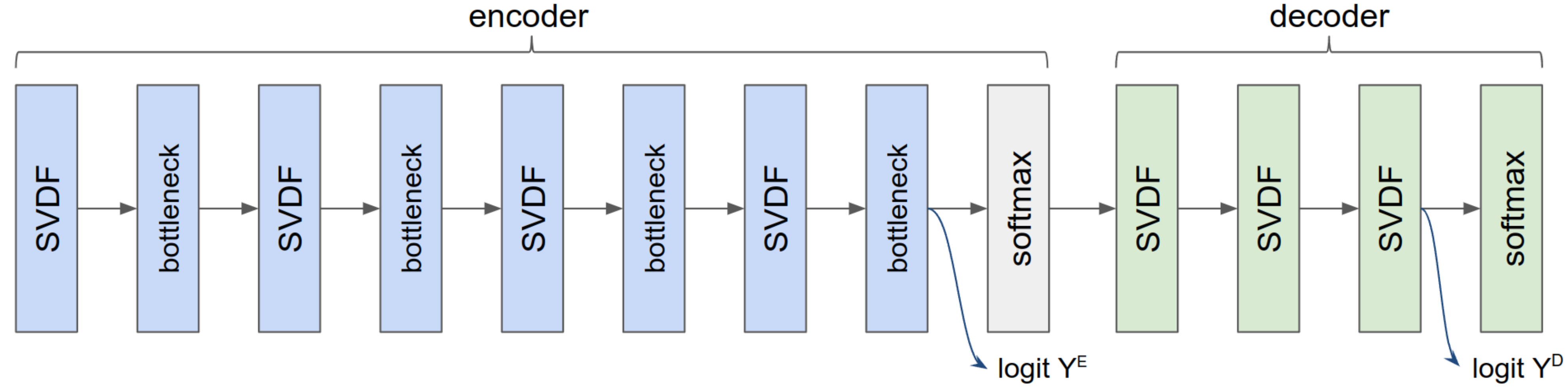}
	\caption{Stacked SVDF with bottleneck model architecture}
	\label{fig:architecture}
\end{figure}

The two-stage model architecture (Fig.~\ref{fig:architecture})
from~\cite{MaxPool20, Alvarez2019} was used for both the teacher and student
networks. The model consisted of 7 factored convolution layers (called SVDF
\cite{Alvarez2019}) and 3 bottleneck projection layers, organized into
sequentially-connected encoder and decoder sub-modules. In total, the model
consisted of approximately 320,000 parameters.

The encoder module processed input features $X_t$, a vector of stacked spectral
filter-bank energies. It generated an $N$-dimensional encoder output
$Y^\textrm{E}$ that was trained to encode $N$ phoneme-like sound units relevant
for the keyword task. The decoder module processed the encoder output and
generated a 2-dimensional output $Y^\textrm{D}$ that was trained to predict the
existence of a keyword in the input audio stream. The combined prediction logit
was defined as $Y=[Y^\textrm{E},Y^\textrm{D}]$.

\subsection{Supervised training objective}
\label{sec:supervised}

The teacher model for FL was trained in a supervised manner. Previous works
proposed two types of supervised losses for KWS model training. The first loss
term directly computed cross-entropy between model logits and
labels~\cite{Alvarez2019}. The second loss term computed cross-entropy between
max-pooled logits and labels~\cite{MaxPool20}. Both loss terms had separate
component terms for the encoder and decoder, and weighted combinations of all
terms were used as the final loss.

In this work, a weighted combination of the cross-entropy and max-pooled losses
was used, as the combination conferred additional regularization and
generalization benefits.

\begin{equation}
\label{eq:mlmp-loss}
\mathcal{L}_{\textrm{sup}} = L_{\textrm{CE}}\left (Y(X_t,\theta), c_t  \right ) + \alpha \cdot L_{\textrm{MP}}\left (Y(X_t,\theta), \omega_{\textrm{end}}  \right )
\end{equation}

$Y(X_t,\theta)$ represents the combined encoder and decoder model output given
input $X_t$ and parameter set $\theta$. $L_{\textrm{CE}}$ represents the
end-to-end loss proposed in \cite{Alvarez2019}. The implementation as defined
by eq. (2) in \cite{MaxPool20} was used, where $c_t$ is the per-frame target
label for CE-loss. $L_{\textrm{MP}}$ represents the max-pool loss proposed in
\cite{MaxPool20}, which was defined by eq. (12) in \cite{MaxPool20}.
$\omega_{\textrm{end}}$ represents the end-of-keyword position label for the
max-pool loss. $\alpha$ is a loss-weighting hyper-parameter determined
empirically. Refer to~\cite{MaxPool20, Alvarez2019} for details of
$L_{\textrm{MP}}$ and $L_{\textrm{CE}}$.

\subsection{Semi-supervised training objective}
\label{sec:semisupervised}

Supervised training was not possible for on-device training, since the
``no-peek" nature of the data precluded manual labeling. The distillation
strategy from~\cite{KwsStce21} was used instead for the FL objective. A teacher
model with the architecture from Section~\ref{sec:architecture} was trained on
server-based data using the supervised training objective from
Section~\ref{sec:supervised}. The fully-trained teacher model, with parameter
set $\theta^{\textrm{Teacher}}$, was used to generate soft labels
$Y^{\textrm{Teacher}}(X_t,\theta^{\textrm{Teacher}})$ using the encoder and
decoder logits for each training example $X_t$. A student model with the same
architecture as the teacher was then trained on these soft labels using the
cross-entropy loss function below.

\begin{equation}
\label{eq:stce-loss}
\mathcal{L}_{\textrm{semi-sup}} = L_{\textrm{CE}}\left ( Y(X_t,\theta), Y^{\textrm{Teacher}}(X_t,\theta^{\textrm{Teacher}}) \right )
\end{equation}

\section{Federated client population}
\label{sec:federation}

Federated training data were stored in local encrypted caches on Android devices
with the Google App\footnotemark. While the app has been installed on more than
10 billion devices as of 2022~\cite{agsa}, caching was only enabled if users
explicitly opted-in to ``save audio recordings on this device and help Google
improve speech technologies for everyone".

\footnotetext{\text{https://play.google.com/store/apps/details?id=}\\ \text{com.google.android.googlequicksearchbox}}

A targeted caching policy, based on the principle of data
minimization~\cite{privacywh, gdpr}, was implemented to ensure that snippets
would only be cached if they could directly improve KWS model reliability. Each
client's local cache stored short snippets of voice recordings that were
captured by the client whenever the Assistant activated or nearly activated, as
well as metadata including how a device was configured, how and when an
interaction with the Assistant happened, and whether the activation was
successful~\cite{assist_support}. Near activation events provided examples of
(true and false) rejects for training, while activation events provided examples
of (true and false) accepts. Clients were allowed to cache up to 20
near-activation examples per day and an unlimited number of activation examples.
Training examples were retained in the cache for 63 days, after which they were
automatically deleted.

When these experiments were conducted in March 2022, the median client cache
contained 175 examples. These examples had an average duration of 1.7 seconds,
with a maximum length determined by the size of the audio
buffer~\cite{small_fp_kws} used for keyword spotting on-device.

Since training cache data were ``no peek" by design, federated KWS training
relied on semi-supervised learning techniques. Labels were inferred from a
teacher model as described in Section~\ref{sec:semisupervised}. Additionally,
on-device signals were used to filter data based on confidence in the labels
(see Section~\ref{sec:filtering}).

Clients were only eligible for federated training if they passed certain
criteria. Devices were required to have at least 2GB of memory, the Google App
must have been updated within the past year, and the device must have utilized
the US English language locale. Once these criteria were met, devices were only
eligible for training during periods of time when they were charging, unused,
connected to un-metered networks, and had non-empty training caches. For each
federated computation task, there was an additional restriction that prevented
any single client from contributing more than once per 24 hours.

At the time of these experiments, nearly 300,000 unique client devices checked
in for training per day and completed up to 6,000 rounds of federated training
per day for cohorts of 400 clients per round. Experiments in this paper took
approximately 1 week to fully converge starting from random weight
initialization.

\section{Filtering}
\label{sec:filtering}

\subsection{User feedback features}

To improve the quality of the unlabeled distillation data, examples were
filtered out if the teacher soft labels were likely to be incorrect. Label
uncertainty was estimated using additional ``feedback features". These features,
which were cached with the audio, indicated whether a later stage keyword model
on the server rejected the utterance, and whether the user followed up a few
seconds after the utterance with a potential re-attempt of the keyword. While
these feedback features indicated whether a label might be incorrect, the
features themselves were noisy.

\subsection{Training cache annotation study}

A user study was conducted to measure label and feedback feature accuracy for
training cache examples. Users downloaded an app that played back their own
cached audio snippets and asked them to indicate whether the examples contained
a keyword. In total, 117 participants annotated 11,908 examples.

Heuristics were then identified that selected for utterances with high-accuracy
teacher labels. Features considered for the selection heuristics included
whether the example was accepted by the on-device KWS model running inference at
the time of collection, the score of that KWS inference model, whether an
unrecognized example had a re-attempt, whether the speaker identification model
recognized the example, and whether a more powerful final-pass KWS server model
accepted the example.

\begin{table}[!h]
\caption{Label accuracy as inferred by on-device signals.}
\centering
\begin{tabular}{ccc} \toprule
\multicolumn{1}{c}{\textbf{Condition}}                                                                                        & \multicolumn{1}{l}{\textbf{Label}} & \multicolumn{1}{c}{\textbf{Accuracy}} \\ \hline \midrule
\begin{tabular}[c]{@{}c@{}}Inference KWS model rejected\\ AND No reattempt\end{tabular}                                               & Negative                                    & $0.91 \pm  0.01$                                                      \\
\hline
\begin{tabular}[c]{@{}c@{}}Speaker ID model rejected\\ AND No reattempt\\ AND Inference Score \textgreater 0.96\end{tabular} & Positive                                    & $0.89 \pm  0.04$                                                          \\
\hline
\begin{tabular}[c]{@{}c@{}}Speaker ID model rejected\\ AND Reattempt\end{tabular}                                             & Positive                                    & $0.88 \pm  0.05$                                        \\
\hline
Server Accepted                                                                                                               & Positive                                    & $0.99 \pm  0.005$                                      \\
\bottomrule
\end{tabular}
\label{tab:filtering}
\end{table}

An analysis of the study data identified a set of conditions that provided a
label accuracy above 88\%, as shown in Table~\ref{tab:filtering}.

\subsection{Example filtering}
\label{sec:example_filtering}

Cached examples that fulfilled the high-accuracy selection conditions in
Table~\ref{tab:filtering} were retained for training, and the remaining examples
were filtered out. While this resulted in a 60\% reduction in the amount of
training data, the excluded data were disproportionately likely to have
incorrect teacher pseudo labels.

An alternative usage of label accuracy information would be to adjust teacher
model logits instead of filtering examples. However, the approach is difficult
to implement in a training recipe that uses continuous soft label targets.
Target alignment is challenging, given that feedback signals provide
per-utterance label corrections, whereas training labels are computed on a
per-frame basis.

Experiments were conducted to determine whether one-sided label adjustment was
possible. Utterances were selected in which the teacher identified as containing
a keyword but the label accuracy heuristics indicated otherwise. The teacher
model scores for those utterances were set to zero to indicate the absence of a
keyword. Unfortunately, these experiments did not yield any improvements over
the filtering approach.

\section{Optimization}
\label{sec:optimization}

Three general training settings were studied: pure FL, pure centralized
training, and joint federated-centralized training.

\subsection{Federated training}
\label{sec:optimization_fl}

The basic federated distillation procedure from \cite{FedHotNonIID20} was reused
in this work. The distillation objective described in
Section~\ref{sec:semisupervised} was used for FL, in conjunction with the data
filtering strategy outlined in Section~\ref{sec:filtering}.

Federated optimization used \texttt{FedYogi}~\cite{adaptivefedopt}, a variant of
the \texttt{FedAvg}~\cite{fedlearn} algorithm which uses the adaptive Yogi
optimizer~\cite{yogi} in lieu of averaging for the server optimizer step. At the
beginning of each federated round, a cohort of 400 client devices was formed
from the eligible population of devices. A set of global KWS model weights and a
training plan were sent to the client devices. The clients processed local
training examples in parallel, taking up to 640 steps of SGD or completing 40
local training epochs before uploading updated model weights to the server.
Weight updates from all the clients in the cohort were aggregated to produce a
new weight delta. The weight delta was then treated as a pseudo-gradient and
used to compute a step of adaptive optimization with Yogi. This process was then
repeated for up to 2,000 rounds until the model converged. Across all training
rounds and clients, this procedure amounted to 512M steps of training.

Hyper-parameters were tuned using the simulation procedure described in
Ref.~\cite{FedHotNonIID20}. These included the client training stopping criteria
(40 local epochs or 640 steps), the optimizer configuration (Yogi with
$\beta_1=0.9$, $\beta_2=0.999$, $\epsilon=0.001$), clipping for client weight
updates ($||L||_2 < 0.5$), softmax temperature ($T=0.9$), client learning rate
schedule (linear warm-up for 40 rounds to a maximum of 0.2, followed by
exponential decay by a factor of 0.9 every 1,000 rounds), and server learning
rate schedule (linear warm-up for 40 rounds to a maximum of 0.2, followed by
exponential decay by a factor of 0.1 every 3,000 rounds).

\subsection{Centralized training}
\label{sec:optimization_ct}

The centralized training procedure mostly followed that described in
\cite{MaxPool20}, using the multi-loss training objective from
Equation~\ref{eq:mlmp-loss}. The underlying server-hosted training data
consisted of vendor-collected utterances, audio collected from YouTube videos,
and utterances from logs.

Asynchronous SGD with 200 parallel trainers and 32 parameter servers was used
for optimization. The learning rate was scheduled with 25M steps of linear
warm-up to a maximum value of $1 \times 10^{-5}$, followed by exponential decay
with a half-life of 200M steps. In total, the central training consisted of 500M
steps.

\subsection{Joint federated-centralized training}
\label{sec:optimization_flask}

FL typically has advantages over centralized learning on a server, in that the
training examples gathered \textit{in situ} by edge devices are reflective of
actual inference serving requests. In the case of KWS, cached examples contain
real background noise, variations in speaker voices, and are generally
reflective of the conditions under which users expect KWS models to function.

However, the cached data are still not fully representative of the inference
data distribution, due to constraints or imbalances of \textit{in situ} data
collection. In this specific case of KWS, the data collection policy outlined
in Section~\ref{sec:federation} explicitly prevented the collection of clearly
negative examples. Smart speakers represented another relevant but missing data
domain for KWS: although FL was conducted on phones in these experiments, the
models were expected to perform inference on both phone and speaker devices.

The benefits of FL can be attained while overcoming any remaining train versus
inference distribution skew by mixing additional central (server-based) data
into a FL training process. This affords a ``composite" set of training data
that better matches the inference distribution. Several strategies exist for
performing joint learning~\cite{augenstein2021mixedfl}.

This work leveraged the ``parallel training" strategy. In this algorithm,
500,000 steps of central training were performed in parallel with one round of
FL after starting from the same global model. The model weights from these two
tasks were then combined using a weighted average (central weight = 1.0, FL
weight = 0.1) to form a new global model. Starting from a random model
initialization, this sequence was repeated hundreds of times until the model
converged. The parallel training algorithm parameters, including the averaging
weights and the ratio of central steps to FL rounds, were tuned using
simulations. The central and federated task components followed the
descriptions in Section~\ref{sec:optimization_ct} and
Section~\ref{sec:optimization_fl}, respectively.

\section{Experiments}
\label{sec:experiments}

Inference quality of the KWS models was assessed using offline evaluation sets
and experiments on production Android phones. Comparisons were made for
centrally-trained, federated, and jointly trained models, as well as models
trained using the example filtering technique from
Section~\ref{sec:example_filtering}.

\subsection{Offline evaluations}

Offline evaluation data were used to quantitatively measure model quality.
Numerous data domains were covered, including different device types (phones and
speakers), noise conditions (in-car recordings, background music), and language
locales (US English, British English, etc.). The data also provided a mix of
positive and negative utterances.

The number of false acceptances per hour of audio (FA/h) was computed for each
model using the negative utterances, while the false reject rate per keyword
instance (FRR) was measured using the positive utterances. An area under the
curve (AUC) metric was defined by integrating the FR curve over the FA/h range
of $[0.05, 0.5]$. Lower values were better for this threshold-independent
metric.

\begin{figure}
  \centering
  \includegraphics[width=\columnwidth]{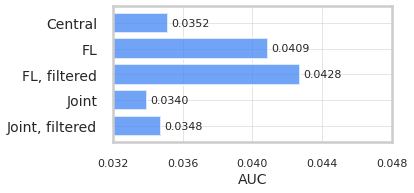}
  \caption{AUC comparisons for different model training techniques. Lower
           values are better.}
  \label{fig:ablation}
\end{figure}

Figure~\ref{fig:ablation} compares AUC measurements for the models. Minimum
values from ensembles of 5 training tasks are shown. Jointly-trained models
yielded the lowest AUC, while FL-only training and example filtering led to
worse AUC.

\subsection{Online A/B experiments}

To assess the actual user-experience of the KWS models, inference experiments
were conducted on a subset of production phones running the Google App. As with
the federated training caches, ground-truth labels were not available to compute
metrics. Instead, proxy false accept ($P_{\textrm{FA}}$) and proxy true accept
($P_{\textrm{TA}}$) metrics were defined based on certain types of Assistant
interactions following probable keywords. While both metrics were correlated
with the ground-truth false accept and true accept rates, the absolute values
of the metrics were not comparable. Therefore, only relative changes in the
proxy metrics ($\Delta P_{\textrm{FA}}$ and $\Delta P_{\textrm{TA}}$) were
compared.

%

Results of production experiments are shown in Table~\ref{tab:live_exp}. The
federated and jointly-trained models significantly improved
$\Delta P_{\textrm{FA}}$ over the control as well as the centrally-trained arm,
while minimizing changes to $\Delta P_{\textrm{TA}}$.

\begin{table}[!h]
  \caption{Relative changes in live A/B experiment metrics with respect to a
           centrally-trained control model. Results are presented using the
           95\% confidence interval for participating devices. Lower
           $\Delta P_{\textrm{FA}}$ is better, while higher
           $\Delta P_{\textrm{TA}}$ is better.}
  \centering
  \begin{tabular}{lcc} \toprule
    Training task   & $\Delta P_{\textrm{FA}}$ [\%]  & $\Delta P_{\textrm{TA}}$ [\%] \\ \midrule
    Central                   & $[-0.91, +11]$      & $[+1.5, +23]$    \\
    FL, averaged              & $[-9.9, -7.4]$      & $[-7.4, -0.43]$  \\
    FL, filtered              & $[-11, -8.6]$       & $[-5.8, +1.0]$   \\
    Joint, averaged, filtered & $[-23, -19]$        & $[-10, +4.1]$    \\
  \bottomrule
  \end{tabular}
  \label{tab:live_exp}
\end{table}

Some details should be noted. The ``central" experiment arm was stopped early.
This resulted in larger statistical uncertainty, and could have introduced
additional systematic uncertainties not covered by the quoted 95\% CI. The
``Joint, averaged, filtered" experiment was conducted a month later than the
others, so the confidence intervals were adjusted to account for changes to the
control model. Finally, the ``averaged" models were post-processed using
checkpoint averaging~\cite{utans96}.

The differences between offline and live experiments were significant. Offline
AUC was likely biased in favor of centrally-trained models due to domain
overlap between the training and testing data partitions. FL and filtering both
under-performed on AUC but improved $\Delta P_{\textrm{FA}}$ and $\Delta
P_{\textrm{TA}}$.

\section{Conclusions}
\label{sec:conclusions}

We trained a keyword spotting model from scratch using federated learning and
reduced mis-triggers for real users when compared with models that were
centrally-trained. Improvements to the canonical FL paradigm were introduced:
example filtering based on user-feedback signals improved label quality for
distillation, and jointly training on federated and central data compensated for
data domains that were absent in the on-device training caches. To our
knowledge, this was the first user-facing launch of an FL-trained model for
speech inference, the first FL model to employ contextual user-feedback signals
to correct semi-supervised training labels, and the first production utilization
of joint federated-centralized training.
\section{Acknowledgements}
\label{sec:acknowledgements}

The authors would like to acknowledge the contributions of Chen Chen, Katie
Knister, Tamar Lucassen, Uli Sachs, Mohammadinamul Sheik, Eric Tung, Lingfeng
Xu, Dan Zivkovic, and colleagues on the Google Research and Speech teams.

\bibliographystyle{IEEEtran}

\bibliography{paper}

\end{document}